\documentclass{article}

\usepackage{microtype}
\usepackage{graphicx}
\usepackage{booktabs}
\usepackage{amsmath}
\usepackage{tikz}
\usetikzlibrary{positioning,arrows.meta,decorations.pathreplacing}
\usepackage{hyperref}

\usepackage[accepted]{mlsys2026}   % remove [accepted] for blind submission

\newcommand{\sys}{FusionML}
\newcommand{\x}[1]{$#1\times$}

\definecolor{oiblue}{HTML}{0072B2}
\definecolor{oiorange}{HTML}{E69F00}
\definecolor{oiverm}{HTML}{D55E00}
\definecolor{oisky}{HTML}{56B4E9}
\definecolor{oigray}{HTML}{777777}

\mlsystitlerunning{\sys{}: Prefill-Scoped CPU+GPU Co-Execution on Apple Silicon}

\begin{document}

\twocolumn[
\mlsystitle{\sys{}: Prefill, Not Decode - Mechanism and Boundaries of
CPU+GPU Co-Execution on Unified-Memory Apple Silicon}

\centerline{\large\bf Om Mohite}
\vskip 4pt
\centerline{Independent Researcher, Mumbai, India}

\mlsyskeywords{Machine Learning, Systems, Apple Silicon, Unified Memory,
Heterogeneous Computing, LLM Inference, Prefill}

\vskip 0.3in
]

\begin{abstract}
Apple-Silicon SoCs share CPU, GPU, and Neural Engine over one unified
memory system, raising the question of whether transformer inference can
be accelerated by splitting single operators across units. Prior
attempts, including our own, failed or produced precision-confounded
wins. We identify the cause: MLX's lazy-graph scheduler
\emph{serializes} cross-stream work whenever a CPU-stream operation
consumes an unmaterialized GPU result inside one evaluation graph, so a
row-split matmul that runs \x{1.38} faster with materialized inputs runs
\x{0.66} slower than GPU-only inside a lazy graph; an eager
materialization boundary restores concurrency (\x{1.34}). \sys{}
implements a per-layer, contention-aware CPU+GPU row split for
transformer prefill built on this fix. Evaluated across five chips and
three Apple-Silicon generations, community-replicated, the split
accelerates Llama-shaped decoder-block prefill by \x{1.15}--\x{1.38},
unchanged at full 32-block depth, and reaches \x{1.18}--\x{1.25} faster
time-to-first-token on a real Qwen2.5-7B checkpoint served through stock
MLX-LM, with token-identical outputs and unchanged decode throughput. We
characterize the boundaries equally carefully: decode cannot benefit,
bound by shared bandwidth co-execution does not add; precision-matched
training loses \x{0.86}--\x{0.97} on all five chips; ANE dispatch
overhead excludes it at layer granularity; and a no-regression runtime
gate becomes self-defeating under memory pressure, where probing an
alternative mode evicts the active mode's working set. Code, raw
results, and generation transcripts are released.
\end{abstract}

\section{Introduction}

Unified-memory SoCs are now the dominant local-inference platform, and Apple
Silicon is their most widely deployed instance. Because CPU, GPU, and Neural
Engine (ANE) address the same physical memory, work can in principle be split
\emph{within} a single operator (rows of one matmul computed simultaneously
on CPU and GPU) without any copy. Whether this helps in practice has
remained unclear: published speedups in this space are frequently confounded
(FP16 systems compared against FP32 baselines), and naive implementations
lose outright.

This paper reports what we believe is the first mechanism-level account of
when intra-operator CPU+GPU co-execution helps transformer inference on this
platform, verified end-to-end and replicated by community contributors on
five chips. Our contributions:

\begin{enumerate}
\item \textbf{Mechanism.} We show that lazy-graph evaluation is the hidden
serializer: MLX runs CPU-stream and GPU-stream work concurrently only when
the CPU operation's inputs are already materialized. A minimal probe
(Section~\ref{sec:mechanism}, Figure~\ref{fig:timeline}) makes the effect
quantitative (\x{1.38}~ready~/ \x{0.66}~lazy~/ \x{1.34}~eager-boundary) and
explains a history of failed co-execution attempts, including why an eager
per-layer Swift implementation succeeded where lazy Python graphs failed.
\item \textbf{Design.} \sys{}: a per-layer row-split for prefill
(Figure~\ref{fig:arch}) with (i) eager materialization boundaries, (ii)
per-shape ratios calibrated \emph{under contention} rather than from
solo-unit rates, and (iii) a runtime gate (probation + periodic re-probing
+ hysteresis) that holds a no-regression floor by construction, since the
baseline execution mode is itself one of the gate's arms
(Section~\ref{sec:design}).
\item \textbf{Fair, replicated evaluation.} Under a strict protocol
(Section~\ref{sec:setup}): \x{1.15}--\x{1.38} Llama-shape prefill wins on
all five chips tested (M1, M2, M3~Pro, M4, M4~Pro); unchanged at full
32-block depth; \x{1.18}--\x{1.25} TTFT on a real 7B checkpoint through
stock MLX-LM with token-identical outputs and neutral decode
(Section~\ref{sec:eval}).
\item \textbf{Boundaries, measured.} Decode is structurally out of scope on
shared bandwidth; precision-matched training loses on every chip; ANE
dispatch overhead and three (since OS-fixed) crash conditions exclude the
Neural Engine at layer granularity; and we quantify an \emph{observer
effect}: under memory pressure, a probing scheduler's measurements of
alternative modes are biased $\sim$\x{1.4} by the probe's own working-set
eviction, with $\sim$\x{1.7} reload cost on the next call
(Section~\ref{sec:boundaries}).
\end{enumerate}

We highlight a methodological stance: every headline number in this paper is
measured against a \emph{precision-matched} baseline executed adjacently on
the same machine, and several of our own earlier ``wins'' are retired in
Section~\ref{sec:boundaries} as FP16-vs-FP32 artifacts. Raw result files,
including power source, battery level, thermal notes, and free memory at run
start, accompany every experiment in the repository.

\section{Background and Related Work}

\textbf{Apple Silicon and MLX.} Apple's M-series SoCs share one LPDDR memory
system across CPU (with AMX matrix units reachable via Accelerate), GPU
(Metal), and ANE (reachable only through CoreML). MLX~\cite{mlx2023} is
Apple's array framework for this platform: lazily evaluated, with a
\texttt{stream} abstraction that places operations on CPU or GPU. MLX-LM is
its de-facto LLM serving layer, our end-to-end baseline.

\textbf{Heterogeneous mobile inference.} CoDL~\cite{codl2022} co-executes CNN
operators across CPU and GPU on smartphones and reports the same class of
contention effects we measure (shared-bandwidth degradation of per-unit
throughput); we extend the contention-aware view to LLM prefill on Apple
Silicon and add the lazy-graph serialization mechanism, which has no analogue
in eager mobile stacks. Phase-splitting systems such as
Splitwise~\cite{splitwise2024} and chunked-prefill schedulers such as
Sarathi~\cite{sarathi2023} exploit the prefill/decode asymmetry at cluster
scale; we exploit the same asymmetry \emph{within one SoC}.

\textbf{Kernel- and block-level acceleration.} FlashAttention~\cite{dao2022}
and successors accelerate attention itself; our split leaves attention on the
GPU untouched and is complementary. Speculative
decoding~\cite{leviathan2023,chen2023} accelerates decode with a draft model;
we argue in Section~\ref{sec:boundaries} that a CPU/ANE-hosted draft is the
natural way to use idle units during decode, precisely because
intra-operator splitting cannot help there.

\section{The Serialization Mechanism}
\label{sec:mechanism}

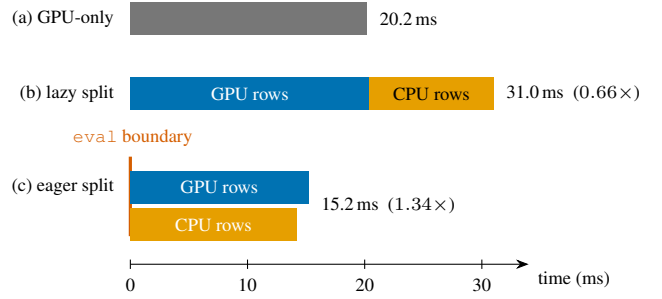
\begin{figure}[t]
\centering
\begin{tikzpicture}[xscale=0.155, yscale=0.62, font=\scriptsize]
% time axis in ms; measured on M1 (2048x1600 @ 1600x6400 fp16, rho=0.35)
\foreach \y/\label in {0/{(a) GPU-only}, -1.6/{(b) lazy split}, -3.6/{(c) eager split}}
  \node[anchor=east] at (-0.6,\y-0.35) {\label};
% (a) GPU-only: 20.2ms
\fill[oigray]  (0,0) rectangle (20.2,-0.7);
\node[anchor=west] at (20.5,-0.35) {20.2\,ms};
% (b) lazy: serialized gpu 20.4 then cpu ~10.6 -> wall 31.0
\fill[oiblue]  (0,-1.6)  rectangle (20.4,-2.3);
\fill[oiorange](20.4,-1.6) rectangle (31.0,-2.3);
\node[anchor=west] at (31.3,-1.95) {31.0\,ms \ (\x{0.66})};
\node[white] at (10.2,-1.95) {GPU rows};
\node at (25.7,-1.95) {CPU rows};
% (c) eager: concurrent, wall 15.2
\draw[very thick, oiverm] (0,-3.3) -- (0,-5.0);
\node[oiverm, anchor=south] at (0.2,-3.28) {\texttt{eval} boundary};
\fill[oiblue]  (0,-3.6) rectangle (15.2,-4.3);
\fill[oiorange](0,-4.4) rectangle (14.2,-5.1);
\node[white] at (7.6,-3.95) {GPU rows};
\node[white] at (7.1,-4.75) {CPU rows};
\node[anchor=west] at (15.5,-4.35) {15.2\,ms \ (\x{1.34})};
% axis
\draw[-{Stealth}] (0,-5.6) -- (34,-5.6) node[anchor=north west, yshift=1pt] {time (ms)};
\foreach \t in {0,10,20,30} \draw (\t,-5.5) -- (\t,-5.7) node[below] {\t};
\end{tikzpicture}
\caption{Execution timeline of one row-split matmul
($2048{\times}1600 \times 1600{\times}6400$, FP16, 35\% CPU rows, M1;
measured wall times). Inside one lazy graph~(b) MLX serializes the streams
and the split \emph{loses}; an eager materialization boundary~(c) restores
concurrency.}
\label{fig:timeline}
\end{figure}

Row-splitting one matmul across \texttt{mx.cpu} and \texttt{mx.gpu} streams
is straightforward; making it concurrent is not.
Figure~\ref{fig:timeline} shows the probe that isolates the effect. When
both stream branches read an already-materialized array, MLX executes them
concurrently and the split delivers the expected win (\x{1.38}). When the
same split consumes the output of a GPU operation \emph{within the same
evaluation graph} (the situation of every layer after the first in any
real network), execution serializes and the split is strictly worse than
doing nothing (\x{0.66}). An explicit materialization boundary
(\texttt{mx.eval} on the input) before the split restores concurrency at
negligible cost (\x{1.34}).

This single scheduling property explains a striking prior split within our
own project: an eager, per-layer Swift implementation showed consistent
co-execution wins while every lazy-graph Python variant, including a
coarse whole-FFN split whose CPU portion waited on attention, lost. The
lesson generalizes: \textbf{lazy-graph frameworks cannot exploit
intra-operator heterogeneous parallelism without eager boundaries}, an
API-level consideration for any unified-memory array framework.

\section{\sys{} Design}
\label{sec:design}

\begin{figure}[t]
\centering
\includegraphics[width=\columnwidth]{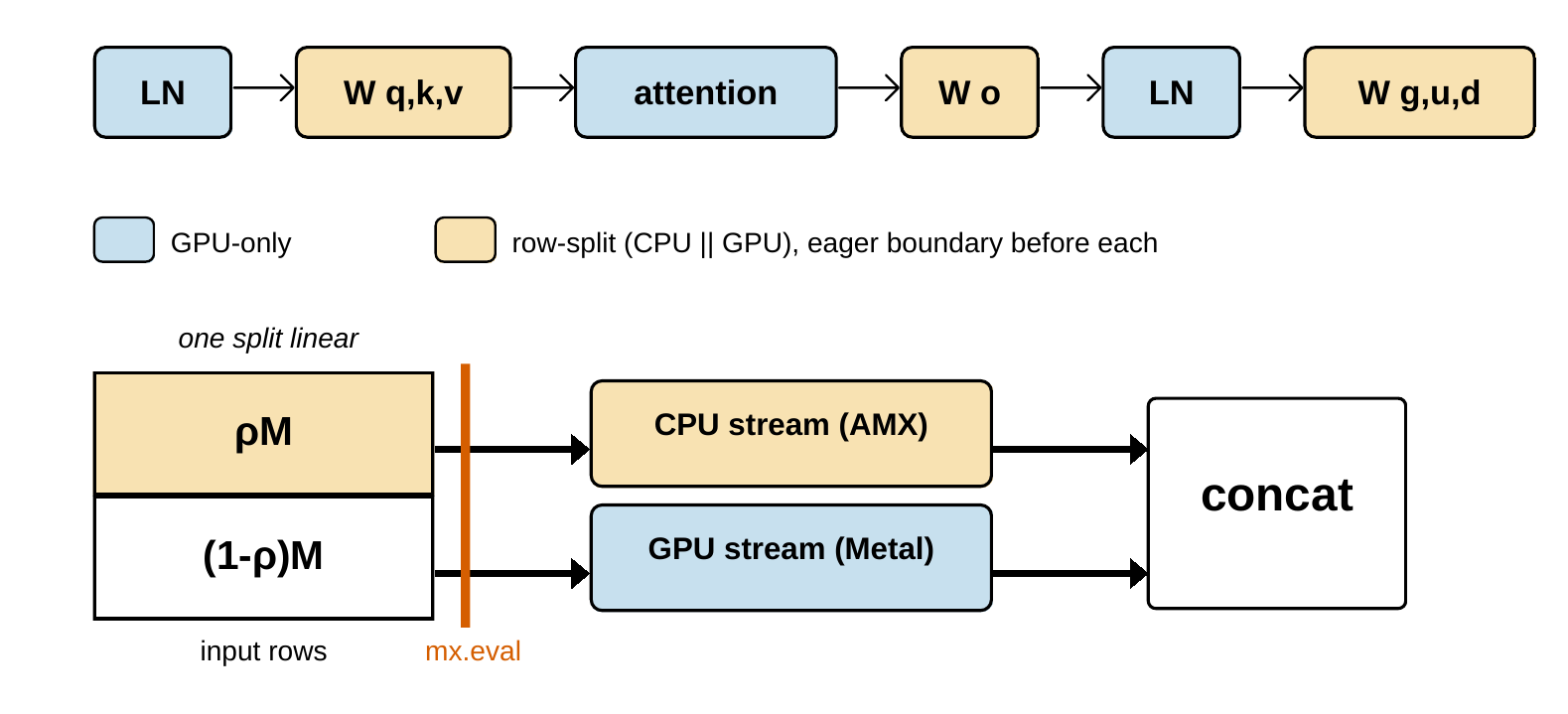}
\caption{\sys{} per-layer prefill split. Every linear layer is row-split
across concurrent CPU/GPU streams behind an eager boundary; attention,
normalization, and element-wise math stay on the GPU. The CPU share $\rho$
is calibrated per matmul shape \emph{under contention}
($\rho \approx 0.25$--$0.40$ across chips).}
\label{fig:arch}
\end{figure}

\textbf{Per-layer split.} Every linear-layer matmul in a decoder block
(QKV/output projections; gate/up/down or FC layers) is row-split: the first
$\rho M$ rows execute on the CPU stream, the remainder on the GPU stream,
concatenated afterward (Figure~\ref{fig:arch}). Attention
score/softmax/value math stays on the GPU. Each split is preceded by an
eager boundary per Section~\ref{sec:mechanism}.

\textbf{Contention-aware calibration.} The ratio $\rho$ is selected per
matmul shape by measuring the \emph{actual concurrent split} over a small
candidate grid ($\rho \in \{0, 0.10, \dots, 0.50\}$), not by composing
solo-unit throughputs: under co-execution, GPU throughput itself degrades
(13.2 $\rightarrow$ 15.5\,$\mu$s/row on M1) because the units share
bandwidth, so solo rates systematically overstate the optimal CPU share.

\textbf{No-regression gate.} A runtime controller treats \{split,
compiled-baseline, eager-baseline\} as competing execution modes: a short
probation picks the fastest, and the inactive modes are re-probed
periodically (every 10th call, exponential-moving-average tracking, 3\%
switching hysteresis). Because the stock execution path is itself an arm,
the floor is the baseline minus bounded probe overhead ($\le$3\% on the
smallest cells we measure, $<$1\% typically). The eager-baseline arm is
necessary: \texttt{mx.compile}d execution (\sys's original default)
itself loses $\sim$6\% to eager MLX at sequence length 8192, so a gate over
\{split, compiled\} alone has no true floor.

\section{Evaluation}
\label{sec:eval}

\subsection{Setup and Fairness Protocol}
\label{sec:setup}

\textbf{Machines.} Five chips across three generations, four of them run by
community contributors from the public repository: M1 (8\,GB, fanless
MacBook Air), M2 (8\,GB), M3~Pro (18\,GB), M4 (24\,GB), M4~Pro (24\,GB); all
on mains power.

\textbf{Protocol.} All comparisons are precision-matched (FP16 vs.\ FP16)
and measured \emph{adjacently} (baseline and treatment interleaved in the
same session, one process at a time, cooldowns between subprocesses).
$n{=}50$ timed runs after 10 warmups per cell (full-depth: $n{=}10$/3).
Every result JSON embeds hardware identity, power source, battery level,
thermal notes, and free memory at start; one contributed run that violated
the memory requirement (a 15.6\,GB model resident in 1.8\,GB) was detected
from these fields and excluded, and the harness now refuses such runs.
Split outputs are verified against GPU-only execution (max relative error
$\le 3{\times}10^{-3}$ at FP16 accumulation, $\le 3.9{\times}10^{-3}$
through 32 blocks).

\subsection{Block-Level Prefill Across Five Chips}

\begin{figure*}[t]
\centering
\includegraphics[width=0.78\textwidth]{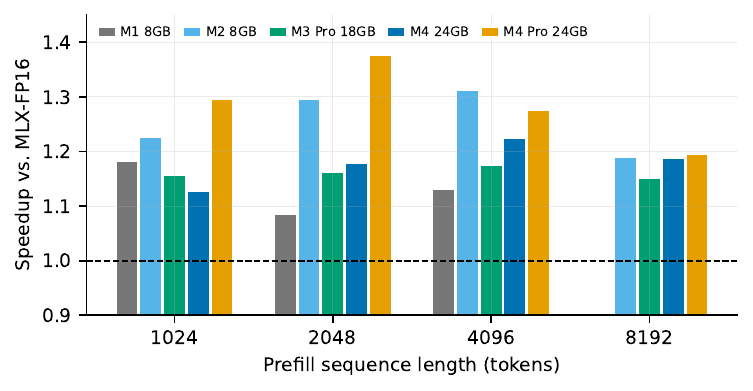}
\caption{Llama-3-8B-geometry~\citep{llama3} decoder-block prefill: per-layer split speedup
over precision-matched MLX-FP16, five chips, sequence length 1024--8192.
Every actively-cooled chip wins at every length; the fanless 8\,GB M1 fades
at 8192 under memory pressure and thermal throttling.}
\label{fig:fivechip}
\end{figure*}

Figure~\ref{fig:fivechip}: the split wins \x{1.15}--\x{1.38} on Llama-shaped
blocks on every chip with active cooling, replicated across independent
sessions ($\pm 0.06$ across two sessions on M4). Two honest qualifications.
First, GPT-2-shaped blocks ($d{=}1600$) win less (\x{1.02}--\x{1.18}) and
are noise-sensitive on 8\,GB machines at short sequence lengths: short
CPU chunks are hypersensitive to scheduler and thermal state. Second, we
predicted win magnitude would track the chip's CPU:GPU core ratio; the
prediction \emph{partially failed} (M2 outperforms M4; M3~Pro trails both),
and we report magnitude as chip-dependent. The robust cross-chip claim is
the consistency of the win, not a single-variable law.

\subsection{Full Model Depth}

\begin{figure}[t]
\centering
\includegraphics[width=0.9\columnwidth]{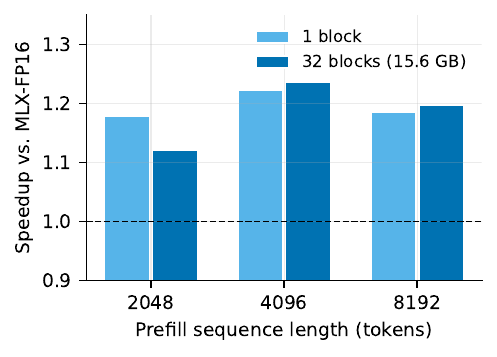}
\caption{Single block vs.\ 32 stacked blocks (15.6\,GB FP16 weights), M4
24\,GB. Depth does not erode the win.}
\label{fig:fulldepth}
\end{figure}

Stacking 32 blocks (15.6\,GB of weights, per-block eager boundaries) leaves
the win essentially unchanged (Figure~\ref{fig:fulldepth}:
\x{1.12}/\x{1.24}/\x{1.20} at 2048/4096/8192 vs.\
\x{1.17}/\x{1.25}/\x{1.17} single-block). Per-layer synchronization cost
does not compound with depth, and correctness holds through the full stack.

\subsection{End-to-End: Real Checkpoint, Real Runner}

\begin{figure}[t]
\centering
\includegraphics[width=\columnwidth]{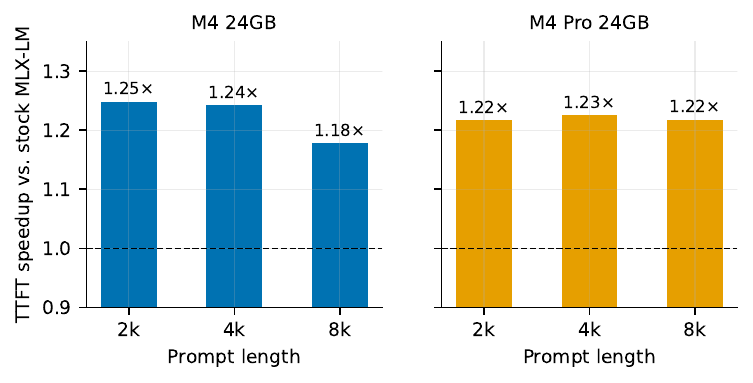}
\caption{Time-to-first-token speedup for Qwen2.5-7B-Instruct (bf16) served
through stock MLX-LM with \sys's split patched into \texttt{nn.Linear}
(prefill-scale calls only; untuned fixed $\rho{=}0.30$). Greedy decoding;
decode throughput unchanged (7.1$\rightarrow$7.1 tok/s on M4,
13.4$\rightarrow$13.5 on M4 Pro).}
\label{fig:ttft}
\end{figure}

We patch MLX-LM's \texttt{nn.Linear} so that large-row (prefill) calls use
the split and decode calls (single-row) fall through untouched, and serve
Qwen2.5-7B-Instruct-bf16~\citep{qwen2024} with greedy decoding. TTFT improves
\x{1.18}--\x{1.25} across 2k/4k/8k-token prompts on both 24\,GB machines
(Figure~\ref{fig:ttft}); decode throughput is exactly neutral, as designed.
Outputs are token-identical to stock in 5 of 6 configurations; in the
sixth, generation diverged at one near-tie token after 289 identical
characters, due to FP16 reduction-order sensitivity, with both completions
coherent. Verbatim prompt/response transcripts for every run ship with the
repository. TTFT is the latency that dominates prefill-heavy workloads
(RAG, document QA, long-prompt/short-answer), which is precisely the regime
this mechanism targets.

\subsection{The Gate, and the Observer Effect}
\label{sec:gate}

In-regime, the gate behaves as constructed: on M4 it holds $\ge$\x{1.0} in
every cell, riding the split where it wins and standing down to baseline
where it does not (e.g., \x{0.98} worst-case on a 15\,ms cell where probe
overhead is proportionally largest). Out of regime it taught us something
better than success. On the 8\,GB M1 at sequence length 8192, where
three execution modes' buffers cannot co-reside, the gate correctly
selected the fastest mode \emph{by its own measurements} and still lost:
each probe of an inactive mode evicted the active mode's working set,
inflating the probe's reading $\sim$\x{1.4} above that mode's true
fresh-process latency and imposing $\sim$\x{1.7} reload cost on the next
active call. A periodically-probing scheduler under memory pressure cannot
observe its alternatives without damaging both the measurement and the
system. The practical rule we adopt: continuous probing only where working
sets co-fit; probation-then-lock with drift-triggered (not scheduled)
re-probing otherwise.

\section{Boundaries: Where Co-Execution Cannot Help}
\label{sec:boundaries}

\textbf{Decode.} Each decoded token streams the full weight set through the
memory system once; decode is bandwidth-bound, and unified memory means the
bandwidth is \emph{shared}. Co-execution adds compute, not bandwidth, so no
intra-operator split can accelerate decode on this architecture, confirmed
by our decode-neutral end-to-end measurements. The idle-unit
opportunity during decode is speculative decoding with a CPU/ANE-hosted
draft model, which spends idle \emph{compute} to save \emph{bandwidth per
accepted token}; we leave it to future work.

\textbf{Training (a retired claim).} \sys's earlier training ``wins''
(\x{1.13}--\x{1.35}) were measured against FP32 baselines while computing
in FP16. Precision-matched, single-block training loses on \emph{all five
chips} (\x{0.86}--\x{0.97}), with finite losses verified. We report this as
the strongest honest negative in the paper and a caution for the
literature: of the headline comparisons we began with, every one that mixed
precisions reversed or vanished when matched.

\textbf{Neural Engine.} ANE is reachable only through CoreML, whose
dispatch costs $\sim$20--24\,ms fixed plus $\sim$7\,$\mu$s/row regardless
of chip generation (measured M1 through M4~Pro), useless at layer
granularity where matmuls take 2--20\,ms, though it amortizes at batch
scale, where our contention-aware three-way search reached
\x{1.61}--\x{1.79} burst (\x{1.16}--\x{1.29} sustained) on raw FFN-shaped
matmuls. Three deterministic CoreML/MLX coexistence crashes we isolated on
macOS~25.x (e.g., loading a second distinct compiled shape segfaults the
process) are fixed in macOS~26.3, verified by rerunning our minimal
repros on two machines, but the dispatch overhead stands, and ANE
weight-baking costs 1.4\% relative error on routed rows.

\textbf{Workload shape.} Co-execution is not universal even at batch scale:
a blocked Cholesky whose trailing update is SYRK-shaped never beat CPU-only
Accelerate BLAS (\x{0.90}--\x{0.95}): when a single unit's specialized
path already saturates the bottleneck resource, adding units only adds
contention.

\textbf{Thermal envelope.} On the one passively-cooled machine, sustained
load reduces every speedup (burst \x{1.79} $\rightarrow$ sustained
\x{1.16}--\x{1.29} on raw matmuls) and small-shape results become
session-dependent; we therefore report cold/hot states separately and treat
fanless chassis as the motivating case for runtime adaptivity rather than a
reliable benchmark platform.

\section{Limitations}

Our transformer measurements use decoder-block geometries with synthetic
weights for controlled experiments (latency is weight-value-independent;
correctness is separately verified) and one real 7B checkpoint end-to-end;
broader model coverage (MoE, multimodal) is future work. The mechanism
analysis is specific to MLX's scheduler, though we expect the
eager-boundary requirement to generalize to any lazily-evaluated framework
with device streams. Ratios in the end-to-end experiment are untuned
($\rho{=}0.30$ fixed); per-shape calibration should widen the reported TTFT
wins. Batch serving throughput, where raw-matmul results suggest the
largest headroom, is measured at the operator level but not yet
end-to-end.

\section{Conclusion}

On unified-memory SoCs, intra-operator CPU+GPU co-execution is neither a
myth nor a free lunch: it is a prefill-scoped mechanism, gated by an
eager-boundary scheduling requirement that lazy-graph frameworks violate by
default, bounded by shared bandwidth in decode, by dispatch overhead on the
ANE, and by working-set co-residency for any scheduler that probes its
alternatives. Within its regime it is real, replicated, and free: up to
\x{1.38} block-level and \x{1.25} end-to-end TTFT on stock MLX-LM with
token-identical outputs. We hope the fairness protocol, precision-matched
adjacent baselines, environment-stamped raw results, and retired claims
reported alongside surviving ones, is as useful to the community as the
mechanism itself.

\section*{Acknowledgments}
Community contributors ran the M2, M3~Pro, M4~Pro, and portions of the M4
benchmark suites on their own hardware; they will be named (or credited as
they prefer) in the camera-ready. Code, raw environment-stamped results,
and verbatim generation transcripts:
\url{https://github.com/ommo007/FusionML}.

\bibliography{references}
\bibliographystyle{mlsys2026}

\end{document}